\documentclass[twoside,leqno]{article}

\usepackage[letterpaper]{geometry}
\usepackage{graphicx}
\usepackage{ltexpprt}
\usepackage{hyperref}
\usepackage{amsmath}
\usepackage{amssymb}
\usepackage{cleveref}
\usepackage[algo2e,ruled]{algorithm2e}

\begin{document}

\title{Planar-faced and high-Jacobian two-refinement hexahedral templates}
{
    \author{Hua Tong\thanks{Department of Mechanical Engineering, Carnegie Mellon University.}
    \and Yongjie Jessica Zhang\thanks{Department of Mechanical Engineering, Carnegie Mellon University.}}    
}

\date{}

\maketitle

\fancyfoot[R]{\scriptsize{Copyright \textcopyright\ 2027 by SIAM\\
Unauthorized reproduction of this article is prohibited}}

\begin{abstract}

Automatically generating high-quality conforming hexahedral (hex) meshes from general input boundaries remains a challenging problem, despite the numerical advantages hex elements offer in simulation. Grid-based adaptive refinement, followed by hanging-node removal, is the most robust fully automatic choice. Among two-refinement methods, primal templates have better mesh quality than dual templates but depend on a strong balancing condition that over-refines the grid; neither family guarantees planar quadrilateral (quad) faces. This paper generalizes recent primal three-refinement templates to a two-refinement scheme under a moderate balancing condition, a relaxation of the strong one. 32 special and five fundamental patterns resolve all 144 symmetry-reduced configurations, followed by either a fast greedy algorithm or an integer linear programming (ILP) algorithm that trades extra runtime for a further reduction in element count. The scheme is the first two-refinement method in which all hexes have planar faces, and its meshes feed directly into a recent quality-guaranteed hex mesh reconstruction algorithm. At comparable element counts, it attains a similar Hausdorff ratio (HR) and a higher minimum scaled Jacobian (min SJ) than its three-refinement counterpart.

\end{abstract}

\section{Introduction}
\label{sec:introduction}

The generation of hex meshes represents a core yet challenging problem in computational geometry. Within finite element and isogeometric analysis \cite{zhang2016geometric}, hex elements are favored relative to tetrahedral (tet) elements due to their superior numerical properties, which facilitate enhanced convergence behavior, lower computational overhead, and higher precision across a broad spectrum of physics-based simulations \cite{cifuentes1992performance,benzley1995comparison,wang2004back,wang2021comparison}. Nevertheless, the automated generation of all-hex meshes that conform to intricate geometries while preserving good mesh quality remains difficult despite decades of investigation \cite{owen1998survey,blacker2000meeting,tautges2001generation,shepherd2008hexahedral,zhang2013challenges,schneider2022large}.

A variety of techniques have been proposed to tackle these difficulties. The polycube method establishes an integer grid map that links the input geometry to an axis-aligned polycube domain, thereby inducing a regular grid; however, it can suffer from severe distortion or even failure when complex topologies are involved \cite{gregson2011all,livesu2013polycut,guo2020cut}. Polycube constructions have further been exploited to produce hex control nets and volumetric spline models for isogeometric analysis \cite{zhang2012solid,wang2013trivariate,zhang2013conformal}. Geometric features can be retained through Boolean operations, skeletons, or centroidal Voronoi tessellation \cite{liu2014volumetric,liu2015feature,hu2016centroidal,hu2017surface}, and dedicated software packages have been implemented to interface with commercial solvers such as Abaqus and Ansys LS-DYNA \cite{lai2017integrating,yu2022hexgen}. As an alternative, frame-field-guided parameterization optimizes a frame field that generates an integer grid map from which hexes are extracted \cite{nieser2011cubecover,li2012all,liu2023locally}. Although these methods yield high-quality meshes with well-preserved features, guaranteeing a robust all-hex outcome remains a major challenge. Tet-hex hybrid methods mitigate this robustness issue by permitting a small portion of non-hex elements, typically merging tets into hexes via local heuristics, centroidal Voronoi tessellation, or field guidance \cite{meshkat2000generating,vyas2009tensor,levy2010p,sokolov2016hexahedral,gao2017robust,livesu2020loopycuts}. However, the presence of non-hex elements demands special treatment in certain numerical solvers.

Of all mesh generation strategies, grid-based methods are the most robust and remain the only fully automatic approaches, with Hexotic \cite{marechal2009advances} being a representative example. Such algorithms generally begin from a Cartesian grid that is adaptively refined until a prescribed criterion is satisfied. Criteria frequently employed include error-sensitive functions \cite{zhang2006adaptive,zhang2010automatic,zhang2013robust,hu2013adaptive}, similarity of normals \cite{ito2009octree}, local thickness \cite{livesu2021optimal,pitzalis2021generalized,tong2024hybridoctree_hex,tong2026element}, and the accuracy of surface approximation \cite{gao2019feature,owen2017template}.

Once adaptive refinement finishes, cells of different levels coexist in the grid, producing hanging nodes that lie on the faces or edges of coarser cells yet are vertices of finer neighbor cells, breaking mesh conformity. Such nodes are eliminated by replacing the affected cells with hanging-node-free hexes. Existing methods fall into two families: primal and dual methods, where primal approaches act directly on grid cells. Within three-refinement schemes, the earliest method is unable to handle concave regions and induces severe over-refinement \cite{schneiders1996octree}. It is improved by introducing templates for concave edges, thereby saving elements \cite{ito2009octree}. The treatment of isolated refinement points is refined and the geometric placement of nodes within the templates is tuned \cite{sun2012adaptive}. The templates are enriched to accommodate certain concave-corner configurations \cite{elsheikh2014consistent}, although templates for all configurations are still missing. Most recently, vertex-based templates that solve all 22 vertex-based configurations and edge-based templates that solve all 144 edge-based configurations are proposed \cite{tong2026element}. By recursively applying local refinements to reduce complex configurations to simple ones covered by a fundamental template set, this method cuts final hex counts by factors of several to several hundred compared to prior three-refinement methods while maintaining relatively high min SJ.

In comparison, primal two-refinement schemes are more desirable due to smoother size transitions. Within this family, the pioneering work cannot handle concave regions either \cite{schneiders2000octree}. It is improved by wrapping transition faces in buffer pillow layers, which enables templates to be applied within concave regions rather than propagating refinement to convex regions \cite{ebeida2011isotropic}. However, the inserted layers still induce extra refinement layers that drastically inflate element counts, and the transition patterns become intricate in the concave regions, degrading mesh quality. This trade-off is alleviated through a hybrid strategy: a buffer pillow layer with lower-quality templates is inserted only when completing the concave region into a convex one would inflate the element count excessively; otherwise, the concave region is completed to convex using higher-quality templates \cite{huang2013incorporating}. A per-cell template approach is proposed to replace the global pillowing of prior methods with a vertex-based algorithm built upon four fundamental templates, thereby localizing the refinement outcome while improving mesh quality \cite{owen2017template}. Its remaining drawback is that refinement may still propagate across up to four layers, so the scheme falls short of the ideal, realized by \cite{tong2026element}, in which the template of a cell is determined solely by the refinement states of its vertices or edges. Finally, none of the above two-refinement methods can guarantee that all elements have planar faces, which is a fundamental requirement in quality-guaranteed algorithms such as \cite{tong2025mchex}.

Dual methods constitute an alternative category. The input primal mesh is transformed so that the dual mesh consists purely of hex elements. These methods offer a promising direction for reducing element counts. Their advantage over primal two-refinement approaches is that they require little to no propagation of transition layers: the dual mesh is guaranteed to be all-hex once every interior vertex of the primal mesh is shared by exactly eight polyhedra. This criterion is more intuitive than that of primal schemes and admits more elegant template designs. The development of dual methods has followed a consistent trend toward relaxing the refinement conditions they impose. The first work combines strong balancing conditions with octree pairing \cite{marechal2009advances}. Transition templates are subsequently optimized to reduce element counts \cite{hu2013adaptive}. A substantial leap forward comes from \cite{livesu2021optimal}, which introduces rotation-symmetric templates that reduce irregular edge valences and relax the strong balancing condition to a weak balancing one while dramatically decreasing element counts. To date, the greatest element savings are achieved by \cite{pitzalis2021generalized}, which generalizes octree pairing into a general pairing strategy solved via ILP, reducing the element count even further.

The final phase of the pipeline projects the boundary vertices onto the surface of the input geometry while preserving mesh quality. This operation adjusts only the pointwise geometric coordinates of the mesh and leaves its topology unchanged. Methods addressing this step include \cite{gao2019feature,tong2024hybridoctree_hex,tong2025fast,tong2026hexopt,protais2026versatile}. Recently, the surface can be deterministically constructed via predefined templates \cite{tong2025mchex}, thereby avoiding heuristic optimization; to guarantee that all elements attain a positive Jacobian, however, the method requires the four vertices of each background hex element face to be coplanar.

As is evident from the foregoing discussion, the element count produced by a grid-based hex-meshing algorithm is determined mostly by the construction of the initial grid and the subsequent removal of hanging nodes. No assumptions are made regarding the motivation behind any preexisting refinement of the initial grid, as such decisions are application-dependent and are typically derived from prior knowledge. Instead, this paper studies hanging-node removal, whose sole objective is to guarantee hex-meshability while minimizing the induced increase in element count. Specifically, the primal two-refinement technique is improved by relaxing the strong balancing condition and introducing a new set of templates that mitigate over-refinement. In addition, it maintains high mesh quality and guarantees that every element has planar faces, thereby ensuring that the resulting meshes are directly applicable to \cite{tong2025mchex}.

The remainder of this paper is organized as follows. Section~\ref{sec:refinementConditions} reviews the refinement conditions employed in prior methods and illustrates how the proposed method relaxes them. Section~\ref{sec:edgeBasedTemplates} introduces the proposed edge-based template system. Section~\ref{sec:elementReduction} introduces a greedy algorithm for reducing the number of elements, as well as an alternative ILP formulation that achieves further reductions at the cost of longer computation time. Section~\ref{sec:results} validates the method experimentally on the dataset of \cite{gao2019feature} and demonstrates the advantages of the proposed templates over their three-refinement counterparts in \cite{tong2026element}. Finally, Section~\ref{sec:conclusionAndFutureWork} concludes the paper with a summary of contributions and directions for future research.

\section{Refinement conditions}
\label{sec:refinementConditions}

This section reviews the refinement conditions adopted in primal three-refinement methods and primal or dual two-refinement methods \cite{tong2026element,owen2017template,pitzalis2021generalized}. By highlighting the strengths and weaknesses of these three approaches, the discussion establishes the rationale for the refinement condition in the proposed method.

Unlike three-refinement, where only the balancing condition needs to be imposed on the 27-tree \cite{schneiders1996octree,ito2009octree,sun2012adaptive,elsheikh2014consistent,tong2026element}, two-refinement necessitates that both pairing and balancing conditions be enforced on the octree \cite{marechal2009advances,ebeida2011isotropic,hu2013adaptive,huang2013incorporating,owen2017template,livesu2021optimal,pitzalis2021generalized}. The pairing condition is required by a fundamental limitation of two-refinement: it cannot independently handle the transition from a \(1\times1\) top face to a \(2\times2\) bottom face while maintaining consistent quad patterns on all four side faces. This configuration involves an odd number of quad faces, which has been proven to preclude any valid hexahedralization of the interior \cite{mitchell1996characterization}.

\begin{figure*}[ht]
\centering
\includegraphics[width=\linewidth]{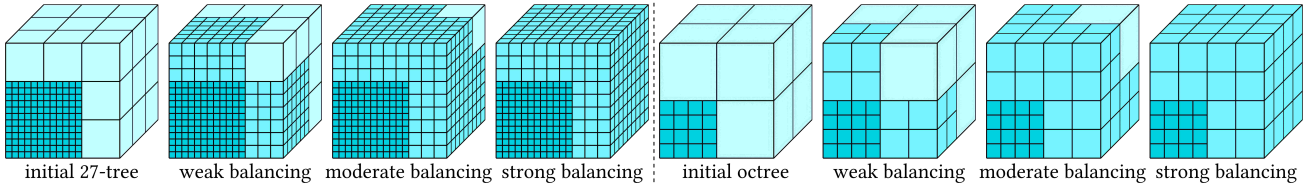}
\vspace{-8mm}
\caption{Refinement results under the initial unbalanced tree, weak, moderate, and strong balancing conditions, applied to a 27-tree (left) and an octree (right). Total cell counts are \(1\,747\), \(2\,059\), \(2\,215\), and \(2\,241\) for the 27-tree (100\%, 118\%, 127\%, and 128\% of the initial count), and \(71\), \(92\), \(113\), and \(120\) for the octree (100\%, 130\%, 159\%, and 169\%). Stronger balancing conditions yield more uniform transitions at the cost of more cells.}
\label{fig:balancingConditions}
\end{figure*}

Three types of balancing conditions have been proposed for refining an initial 27-tree/octree, namely weak, moderate, and strong balancing, as illustrated in \Cref{fig:balancingConditions}. All three conditions bound the tree-level difference between adjacent grid cells, but differ in the definitions of adjacency. The weak balancing condition requires that any two cells sharing a face differ in tree level by at most one \cite{livesu2021optimal}; the moderate condition extends this requirement to cells sharing a face or an edge \cite{tong2026element}; and the strong condition further extends it to cells sharing a face, an edge, or a vertex \cite{marechal2009advances}. In three-refinement, weak balancing has never been adopted; in fact, apart from \cite{tong2026element}, no method can handle all template configurations even under moderate or strong balancing. Nor is weak balancing likely to become attractive in the future, since designing weak-balancing templates for three-refinement would introduce \(1\rightarrow9\) edge transitions and thus produce edges of very high valence. Moderate balancing is adopted to control edge valence \cite{tong2026element}, whereas all earlier methods adopt strong balancing \cite{schneiders1996octree,ito2009octree,sun2012adaptive,elsheikh2014consistent}.

For two-refinement, the pairing condition demands that every transition face be decomposed into multiple \(2\times2\) clusters. The simplest form of octree pairing requires that whenever one child of an octree cell is refined, the other seven children be refined as well \cite{marechal2009advances}. Several methods explicitly adopt this rule \cite{marechal2009advances,ebeida2011isotropic,hu2013adaptive,owen2017template,livesu2021optimal}, while another does not impose it directly but satisfies it implicitly \cite{huang2013incorporating}. More recently, a general pairing scheme has been proposed that employs ILP to reduce the number of resulting octree cells while still guaranteeing the pairing condition \cite{pitzalis2021generalized}. As for the balancing condition, early methods apply the strong balancing condition, either explicitly \cite{schneiders2000octree,marechal2009advances,ebeida2011isotropic,hu2013adaptive} or implicitly \cite{huang2013incorporating,owen2017template}, whereas two recent works relax it to weak balancing \cite{livesu2021optimal,pitzalis2021generalized}. To date, the dual method of \cite{pitzalis2021generalized} is regarded as the most element-saving method. This element-saving property stems from two sources: not only does it adopt the more relaxed weak balancing and general pairing conditions, but the elegant dual construction guarantees an all-hex mesh whenever exactly eight polyhedra share each interior vertex of the primal mesh. Given the simplicity of the primal mesh employed in \cite{pitzalis2021generalized}, it seems unlikely that any primal method could produce fewer hex elements.

This does not mean, however, that primal methods lose value. Their primary advantage over dual methods lies in mesh quality: the min SJ of the primal method of \cite{owen2017template} is 0.329, considerably higher than the 0.0926 of \cite{pitzalis2021generalized}. A second advantage relates to the recent MCHex framework \cite{tong2025mchex}. Rather than the baseline of removing exterior cells and projecting the boundary onto the input geometry, MCHex reconstructs, for each background hex cell, an all-hex mesh conforming to the input geometry via marching cubes and midpoint subdivision, according to whether each of the eight vertices lies inside or outside the domain. This approach offers two theoretical guarantees: the min SJ of every element is provably positive, and the HR from the reconstructed surface to the input geometry converges at a second-order rate with respect to the cell edge length, improving upon the first-order rate of the baseline remove-exterior approach. To guarantee positive min SJ, MCHex requires every background cell to have planar faces. This condition is difficult to satisfy in dual methods but relatively easy in primal methods, since directly controlling vertex positions is far more tractable than controlling the positions of dual vertices inside primal polyhedra. In dual methods, a configuration satisfying the all-planar-face requirement may not even exist. MCHex currently employs the three-refinement templates of \cite{tong2026element}. Building on these observations, this paper develops two-refinement templates that satisfy the all-planar-face requirement while exhibiting lower edge valence; moreover, when fitting the same input geometry with comparable element counts, they achieve a similar HR and a higher min SJ compared to \cite{tong2026element}. Finally, the proposed method is currently restricted to octree pairing; extending it to general pairing is left for future work.

\section{Edge-based templates}
\label{sec:edgeBasedTemplates}

\begin{figure*}
\centering
\includegraphics[width=\linewidth]{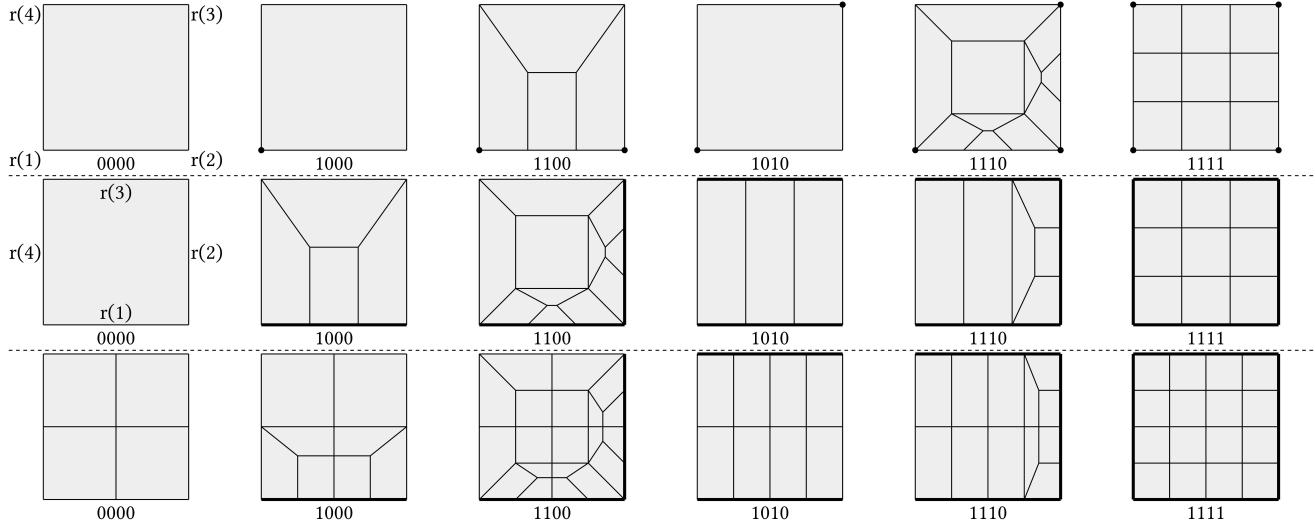}
\vspace{-8mm}
\caption{Face transition templates. The first, second, and third rows show the vertex-based three-refinement, edge-based three-refinement, and edge-based two-refinement templates, respectively. If the \(i\)-th vertex is incident to deeper-level cells, it is marked with a black dot and annotated \(r(i)=1\); otherwise \(r(i)=0\). Likewise, if the \(i\)-th edge is incident to deeper-level cells, it is drawn in bold and annotated \(r(i)=1\); otherwise \(r(i)=0\). It can be verified that, for the vertex-based templates, an edge with endpoint labels \(00\) or \(01\) is unrefined, whereas an edge labeled \(11\) is refined; for the edge-based templates, an edge labeled \(0\) is unrefined, whereas an edge labeled \(1\) is refined. Consequently, the three families are consistent at cell boundaries, so that no hanging nodes arise.}
\label{fig:facePatterns}
\end{figure*}

The templates proposed in this paper generalize those of \cite{tong2026element} to the two-refinement setting. The two-dimensional setting is first introduced. \Cref{fig:facePatterns} demonstrates the three families: the first, second, and third rows show the vertex-based templates of three-refinement, the edge-based templates of three-refinement, and the edge-based templates for two-refinement, respectively. A similar black-dot convention for vertex-based templates appears in Table~I of \cite{ito2009octree}. For a given cell, each vertex is marked with a black dot if any other cell incident to that vertex lies at a deeper tree level than the current cell; otherwise, the vertex is left unmarked. Under this rule, an edge with at most one marked endpoint remains unrefined, whereas an edge with both endpoints marked is split into three segments. Since the tree is required to satisfy the moderate balancing condition, these are the only two patterns that can arise, and an edge is never split into nine segments, so the scheme is self-consistent. The edge-based templates of three-refinement \cite{tong2026element} exploit this property directly: edges that are split into three segments are drawn in bold, while unrefined edges are not.

In this work, two-refinement is built upon octree pairing: each cell whose eight children are all leaves can be treated as a cluster of eight cells. It can be seen that, under moderate balancing, every edge of such a cluster either remains unrefined, consisting of two segments, or is subdivided into four segments; the latter pattern is likewise indicated by bold edges. As can be verified from the third row of \Cref{fig:facePatterns}, all three families of face transition templates are consistent, in the sense that they never leave hanging nodes where two faces meet.

The choice of template family affects element count. In three-refinement, the vertex-based templates form a subset of edge-based templates. The edge-based family therefore affords greater flexibility and potentially yields transitions with fewer elements. The same relationship carries over to two-refinement; consequently, the vertex-based counterpart is omitted and the two-refinement edge-based templates are introduced directly.

\begin{figure*}
\centering
\includegraphics[width=\linewidth]{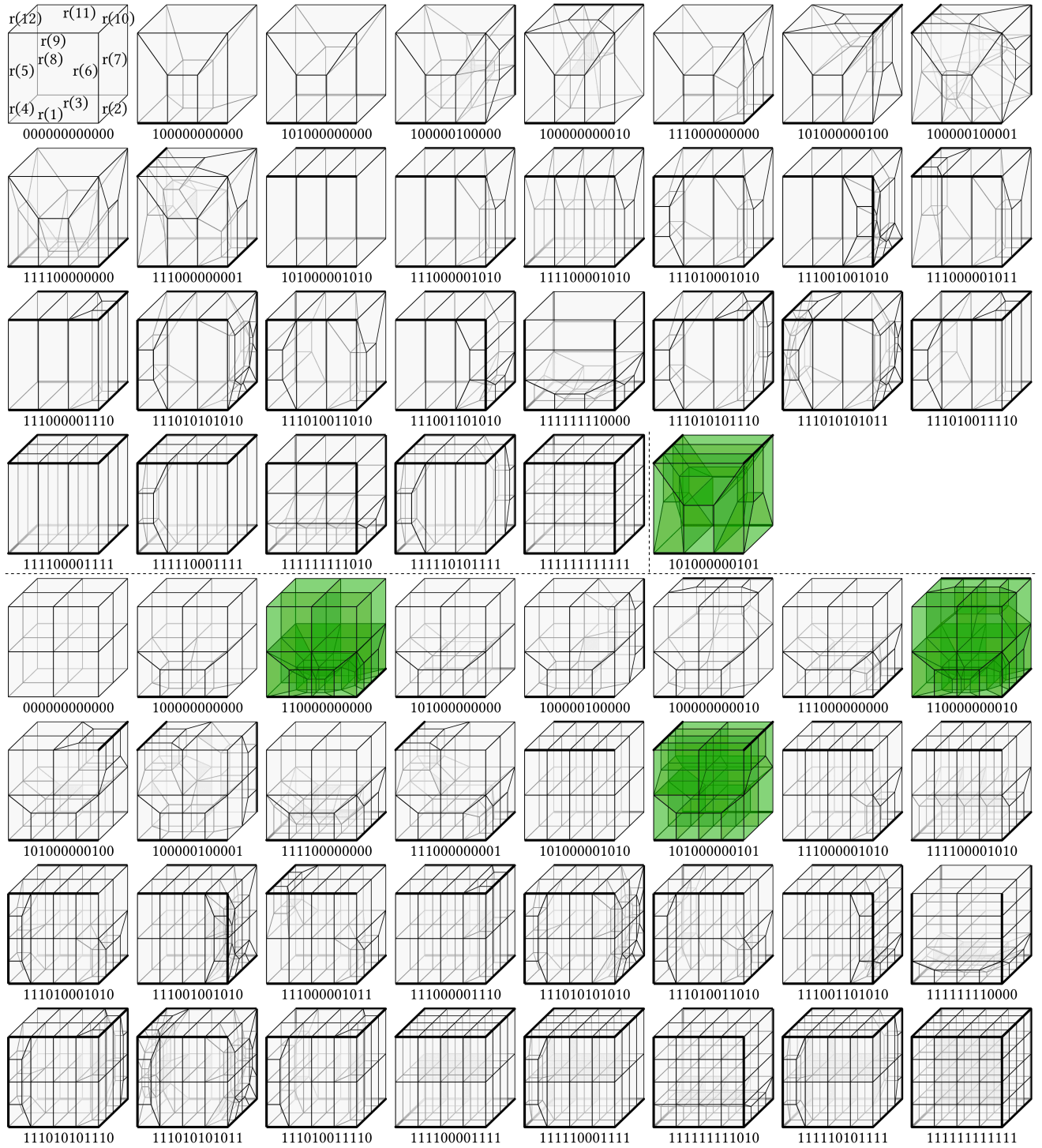}
\vspace{-8mm}
\caption{Special volume transition templates. The templates above the dashed line recall the three-refinement templates of \cite{tong2026element}, while those below the dashed line are the two-refinement templates. For each three-refinement cell and two-refinement cluster, the 12 edges are numbered as indicated in the top-left corner, and each edge is labeled \(r(i)\in\{0,1\}\), where \(r(i)=1\) indicates that the \(i\)-th edge is refined and \(r(i)=0\) that it is unrefined. Green patterns are new relative to \cite{tong2026element}: one appended to the end of the three-refinement family, and three to the two-refinement family. It can be verified that every face of these volume templates matches the corresponding face template of \Cref{fig:facePatterns}; hence both families are consistent at cell boundaries, and no hanging nodes arise.}
\label{fig:specialVolumePatterns}
\end{figure*}

The volume patterns are then introduced. As discussed in \cite{tong2026element}, each cell (or a cluster in the two-refinement setting) has 12 edges, each annotated as 1 if it is refined and 0 otherwise, giving \(2^{12}=4\,096\) possible assignments that reduce to 144 distinct configurations when symmetry is taken into account; in what follows, a configuration is identified by its 12-bit annotation. The top half of \Cref{fig:specialVolumePatterns} shows the three-refinement special volume patterns proposed for 30 of the 144 configurations; the corresponding patterns are displayed explicitly and can be substituted directly. The advantage of such special patterns is that they exploit the symmetry of the configuration and therefore require fewer elements than the fundamental ones. It is thus worthwhile to enlarge the set of special volume patterns so as to reduce the element count as much as possible, while not compromising element quality too much. Balancing these two objectives, the first 29 patterns are obtained in \cite{tong2026element}.

The bottom half of \Cref{fig:specialVolumePatterns} exhibits the two-refinement special volume patterns proposed for 32 of the same 144 configurations. Three green patterns do not appear among the 29 patterns of \cite{tong2026element}: \(110000000000\), \(110000000010\), and \(101000000101\). For the first two configurations, three-refinement has no freedom in the choice of pattern: the only option is to use a local refinement template that decouples the two bold edges of the bottom face (annotated \(1100\)) and treats them separately. Under two-refinement, in contrast, the fundamental pattern would require this template to refine the four upper cells (this will be explained in the discussion of the fundamental volume patterns below), whereas the new special pattern avoids this penetration and saves hexes. Enumeration shows that there are two configurations that benefit from this trick; moreover, since \(110000000000\) corresponds to a concave edge and thus occurs frequently in practice, the saving is particularly valuable. As for \(101000000101\), it can in fact also be treated by a special volume pattern under three-refinement, but this is overlooked in \cite{tong2026element} (the corresponding pattern is shown in the last panel of the top half of \Cref{fig:specialVolumePatterns}); the analogous pattern applies under two-refinement as well.

\begin{figure*}
\centering
\includegraphics[width=\linewidth]{fundamentalVolumePatterns.jpg}
\vspace{-8mm}
\caption{Fundamental volume transition templates comparing three-refinement (left) and two-refinement (right). In both methods, the internal cluster is projected onto two opposite \(1100\) faces (\(1100\mathrm{xxxx}1100\), \(1100\mathrm{xxxx}0110\), \(1100\mathrm{xxxx}0011\), with \(r(i)=\mathrm x\) meaning edge \(i\) is arbitrary) or a single \(1100\) face (\(1100\mathrm{xxxxxxxx}\)) to save elements; the general pattern \(\mathrm{xxxxxxxxxxxx}\) allows no projection. The red patterns show templates applied to \(110000000000\) and \(110000000010\), where two-refinement can save more elements (\Cref{fig:specialVolumePatterns}) but three-refinement cannot.}
\label{fig:fundamentalVolumePatterns}
\end{figure*}

Next, the remaining \(144-32=112\) patterns need to be addressed. As illustrated in \Cref{fig:fundamentalVolumePatterns}, the underlying idea remains consistent with that in \cite{tong2026element}. For each cell in three-refinement or cluster in two-refinement, if there exists a pair of opposite \(1100\) faces, the internal local refinement cluster can be projected onto these two faces, thereby reducing the number of generated elements. By symmetry, this template has three configurations: \(1100\mathrm{xxxx}1100\), \(1100\mathrm{xxxx}0110\), and \(1100\mathrm{xxxx}0011\), where \(r(i)=\mathrm x\) means the \(i\)-th edge state is 0 or 1. Here, transition templates only need to be applied to the four small clusters, excluding those on the top and bottom faces. If no pair of opposite faces is \(1100\), but there exists a \(1100\) face, the internal cluster can be projected onto this face to save elements. By symmetry, this template corresponds to a single configuration \(1100\mathrm{xxxxxxxx}\), requiring templates to be applied to five small clusters, excluding the one on the bottom face. Finally, if no \(1100\) face exists, a standard local refinement is performed. This corresponds to configuration \(\mathrm{xxxxxxxxxxxx}\), where templates must be applied to six small clusters across all faces.

The red patterns in \Cref{fig:fundamentalVolumePatterns} demonstrate that for the \(110000000000\) and \(110000000010\) configurations, three-refinement seems to have no alternative for further element reduction. Meanwhile, two-refinement templates in \Cref{fig:fundamentalVolumePatterns} generate 40 and 52 elements, respectively. However, as illustrated in \Cref{fig:specialVolumePatterns}, the reduction in elements can be achieved by preventing the local refinement of the bottom face from propagating to the four upper cells. This approach requires only 32 and 38 elements for the respective configurations.

\begin{algorithm2e}
\caption{Two-refinement volume transition template generation.}
\label{alg:twoRefinementTemplates}
\KwData{set of clusters \(C\) (each cluster is a \(2\times2\times2\) group of octree cells).}
\KwResult{hex elements in each cluster \(c\in C\).}
\For{cluster \(c\in C\)}{
\For{\(i\gets1\) \KwTo\(12\)}{
\(r(i)\gets\text{edge } i\text{ is refined}\)\;
}
\For{\(i\gets1\) \KwTo\(6\)}{
\(f(i)\gets\text{face } i\text{ forms a } 1100\text{ pattern}\)\;
}
\(usePair\gets f(1)\wedge f(6)\vee f(2)\wedge f(4)\vee f(3)\wedge f(5)\)\;
\(useSingle\gets\neg usePair\wedge(f(1)\vee f(2)\vee f(3)\vee f(4)\vee f(5)\vee f(6))\)\;
\(solved\gets\textsc{False}\)\;
\For{symmetry \(sym\in48\text{ symmetries}\)}{
\If{\(sym(r)\in32\text{ special patterns}\)}{
apply the special template to the transformed cluster \(sym(c)\)\;
\(solved\gets\textsc{True}\)\;
\textbf{break}\;
}
}
\If{\(solved\)}{
\textbf{continue}\;
}
\For{symmetry \(sym\in48\text{ symmetries}\)}{
\(f'\gets sym(f)\)\;
\If{\(usePair\wedge f'(1)\wedge f'(6)\)}{
apply the fundamental template \(1100\mathrm{xxxx}1100\) or \(1100\mathrm{xxxx}0110\) or \(1100\mathrm{xxxx}0011\) to the transformed cluster \(sym(c)\)\;
\textbf{break}\;
}
\ElseIf{\(useSingle\wedge f'(1)\)}{
apply the fundamental template \(1100\mathrm{xxxxxxxx}\) to the transformed cluster \(sym(c)\)\;
\textbf{break}\;
}
\ElseIf{\(\neg usePair\wedge\neg useSingle\)}{
apply the fundamental template \(\mathrm{xxxxxxxxxxxx}\) to the transformed cluster \(sym(c)\)\;
\textbf{break}\;
}
}
}
\Return{hex elements in each cluster \(c\in C\)}\;
\end{algorithm2e}

Finally, all small clusters in the five fundamental templates in \Cref{fig:fundamentalVolumePatterns} have been decoupled such that, of the 12 edges of each small cluster, only the four edges on a single face are flagged with \(0/1\) status. The applicable templates reduce to the six face patterns: \(000000000000\), \(100000000000\), \(110000000000\), \(101000000000\), \(111000000000\), and \(111100000000\). All of them are shown in \Cref{fig:specialVolumePatterns} and can be applied directly. This completes the adaptation of the templates of \cite{tong2026element} to two-refinement. The pseudocode in Algorithm~\ref{alg:twoRefinementTemplates} runs in \(O(\lvert C\rvert)\) time, where \(\lvert C\rvert\) is the number of clusters.

\begin{figure}[ht]
\centering
\includegraphics[width=.5\linewidth]{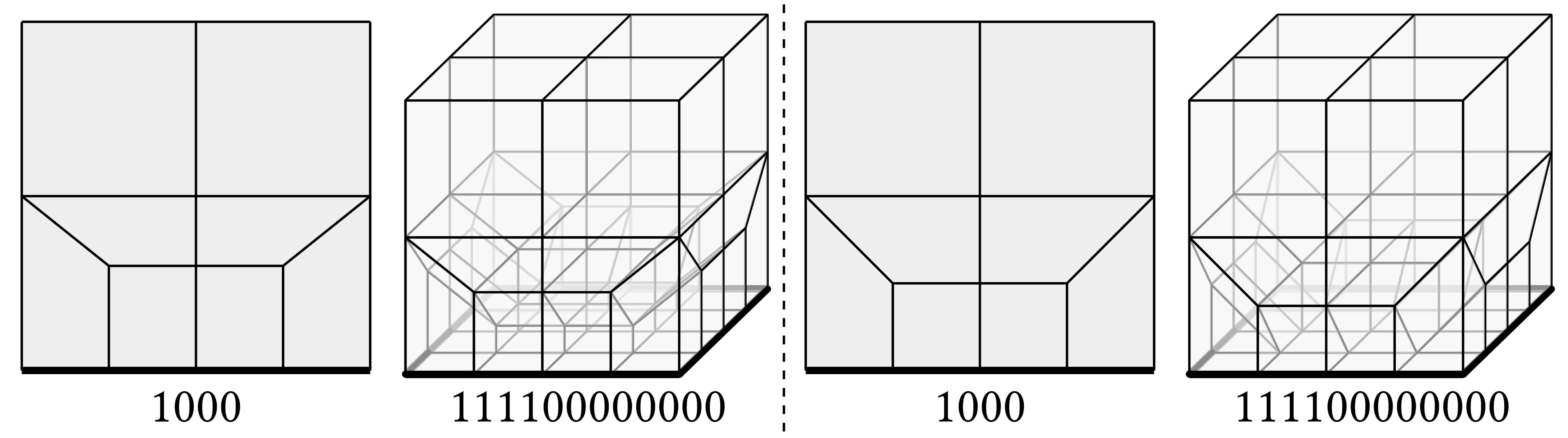}
\vspace{-4mm}
\caption{Necessity of the offset. From left to right: the \(1000\) face template and the \(111100000000\) volume template with the local refinement points shifted by 5\% of the cluster edge length away from the bold refined edge, and the same two templates with the refinement points placed at the center (i.e., without the offset). With the center placement, coplanarity produces degenerate elements.}
\label{fig:offsetIsNecessary}
\end{figure}

\begin{Remark}
Similar to \cite{tong2026element}, the local refinement vertices in \Cref{fig:facePatterns} are shifted by 5\% of the cluster edge length away from the bold refined edge (left in \Cref{fig:offsetIsNecessary}), rather than placed at the center (right in \Cref{fig:offsetIsNecessary}). This offset is required to keep every face planar. If the points are instead placed at the center, the bottom-left hex of the \(111100000000\) template in \Cref{fig:specialVolumePatterns} would be invalid, because coplanarity would force its interior vertex onto the bottom face, precluding a positive Jacobian. Exhaustive enumeration over all templates verifies that every face is planar, providing the first such guarantee among two-refinement methods, and certifies a min SJ of \(0.214\), close to the value in \cite{tong2026element} and higher than that of \cite{pitzalis2021generalized}.
\end{Remark}

\section{Element reduction}
\label{sec:elementReduction}

\begin{figure}
\centering
\includegraphics[width=.5\linewidth]{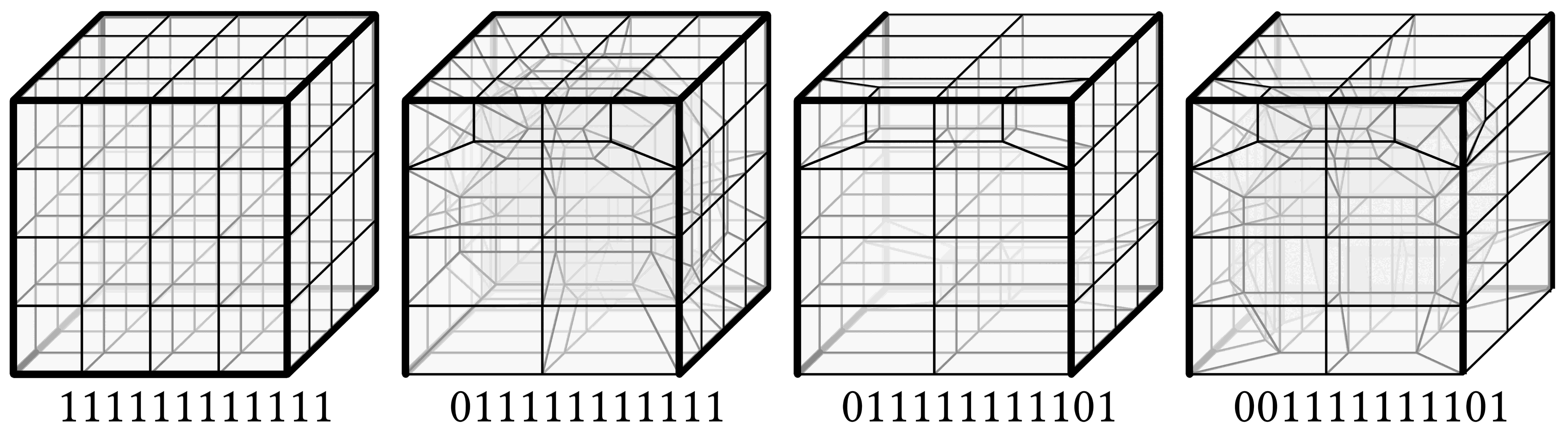}
\vspace{-4mm}
\caption{Having fewer edges annotated as 1 does not necessarily lead to fewer hex elements. From left to right, one edge is switched from 1 to 0 at each step; however, the element counts 64, 156, 44, and 108 do not decrease monotonically.}
\label{fig:fewerRefinementMoreElements}
\end{figure}

As in \cite{tong2026element}, the template set produced by Algorithm~\ref{alg:twoRefinementTemplates} is guaranteed to cover the configurations of its vertex-based counterpart, yet it may contain more hex elements. \Cref{fig:fewerRefinementMoreElements} explains this by showing the volume patterns of four refinement states, whose 12-bit binary encodings are \(111111111111\), \(011111111111\), \(011111111101\), and \(001111111101\), respectively. Moving from left to right, one edge is coarsened from 1 to 0 at each step, so the \(r\) value keeps decreasing; however, the element counts are 64, 156, 44, and 108, which are not monotonic. Consequently, although the template set yields an all-hex solution for every one of the \(4\,096\) states, when the solution of a state contains excessive elements, it can be beneficial to annotate some 0-edges as 1, which reduces the element count. In fact, the vertex-based template set is what the edge-based template set becomes after merging patterns by annotating some 0-edges as 1; this explains why, in \cite{tong2026element}, the hex elements produced directly by the edge-based templates often outnumber those of the vertex-based templates. Therefore, the same greedy algorithm as in \cite{tong2026element} is adopted to save elements; furthermore, an ILP formulation is proposed that takes longer to solve but yields fewer elements. Both methods essentially annotate some 0-edges as 1. However, such an operation changes the states of the clusters adjacent to the current one, and global adjustments may save even more elements; the problem is therefore non-trivial.

\SetKwBlock{initialize}{initialize sets \(C(e)\) and \(E\):}{}
\begin{algorithm2e}[ht]
\caption{Greedy algorithm to reduce elements.}
\label{alg:greedyAlgorithm}
\KwData{set of clusters \(C\); element cost \(\#\mathrm{hex}(x)\).}
\KwResult{some 0-edges annotated as 1.}
\initialize{
\(C(e)\gets\emptyset,E\gets\emptyset\)\;
\For{cluster \(c\in C\)}{
\For{edge \(e\in\) 12 edges of \(c\)}{
\If{edge \(e\) is not refined and refining it would not violate the moderate balancing condition}{
attach cluster \(c\) to set \(C(e)\)\;
attach edge \(e\) to set \(E\)\;
}
}
}
}
\(noIncrease\gets\textsc{True}\)\;
\While{\(noIncrease\)}{
\(noIncrease\gets\textsc{False}\)\;
\For{edge \(e\in E\)}{
\(r_c/r'_c\gets\) encodings of \(c\) before/after refining \(e\)\;
\(n/n'\gets\sum\limits_{c\in C(e)}\#\mathrm{hex}(\sum\limits_{i=1}^{12}2^{12-i}r_c/r'_c(i))\)\;
\If{\(n'\le n\)}{
refine edge \(e\)\;
\(noIncrease\gets\textsc{True}\)\;
\(E\gets E\setminus\{e\}\)\;
}
}
}
\Return{modified edges}\;
\end{algorithm2e}

The greedy algorithm proceeds as follows. A lookup function \(\#\mathrm{hex}:\{0,\dots,4\,095\}\to\mathbb{Z}^+\) is precomputed that records, for each of the \(2^{12}=4\,096\) states, the number of hex elements; here a state is the 12-bit refinement pattern of a cluster, encoded as the decimal value \(\sum\limits_{i=1}^{12}2^{12-i}r(i)\), where \(r(i)\in\{0,1\}\) indicates whether the \(i\)-th edge of the cluster is refined. Next, the candidate edge set \(E\) is built. For each cluster, its 12 edges are traversed and an edge \(e\) is collected into \(E\) if \(e\) is not refined and refining it would not violate the moderate balancing condition. For each \(e\in E\), \(C(e)\), the set of clusters sharing \(e\), is recorded as well; clearly \(\lvert C(e)\rvert\in\{1,2,4\}\). The algorithm then iterates: in each pass, every \(e\in E\) is examined, where \(r_c\) and \(r'_c\) denote the 12-bit states of cluster \(c\) before and after refining \(e\), respectively (refining \(e\) flips its bit from 0 to 1). If refining \(e\) does not increase the total number of elements of the affected clusters, i.e., \(\sum\limits_{c\in C(e)}\#\mathrm{hex}(\sum\limits_{i=1}^{12}2^{12-i}r'_c(i))\le\sum\limits_{c\in C(e)}\#\mathrm{hex}(\sum\limits_{i=1}^{12}2^{12-i}r_c(i))\), \(e\) is refined and removed from \(E\). Since each refinement changes the states of the related clusters, all subsequent checks are based on the updated states. The algorithm terminates when a complete pass refines no edge. The pseudocode is given in Algorithm~\ref{alg:greedyAlgorithm}. It runs in \(O(k\lvert C\rvert)\) time, where \(k\) is the number of greedy passes, typically fewer than 10 in the experiments.

Although Algorithm~\ref{alg:greedyAlgorithm} is fast, it often yields suboptimal solutions: as discussed earlier, global adjustments may lead to fewer elements. The ILP formulation is more element-efficient \cite{pitzalis2021generalized}. As with the greedy approach, the ILP method first collects the set \(E\) of refinable edges, i.e., edges that are unrefined and whose refinement would not violate the moderate balancing condition. This criterion is kept deliberately: since the bottleneck is the solver's runtime, edges that could break the balance are excluded upfront, as admitting them would further inflate the problem size. Even so, incorporating every refinable edge can make the problem prohibitively large and slow the solver. The scale of \(E\) must therefore be restricted; the empirical performance of this restriction is reported in Section~\ref{sec:results}. Specifically, \(E\) is constructed under the following constraint: an edge \(e\) is kept only if at least one cluster incident to \(e\) contains more than four refined edges among its 12 edges; otherwise, \(e\) is discarded. The threshold of four is an empirical value determined by the computation time, and it can be adjusted in practice. Setting the threshold to the minimum \(-1\) keeps all edges in \(E\) (slowest), whereas setting it to the maximum \(11\) leaves \(E\) empty (no optimization). This strategy is merely one feasible heuristic; alternative and potentially superior pruning schemes may exist, but their exploration lies beyond the scope of this paper.

\begin{algorithm2e}
\caption{ILP algorithm to reduce elements.}
\label{alg:ILPAlgorithm}
\KwData{set of clusters \(C\); element cost \(\#\mathrm{hex}(x)\).}
\KwResult{some 0-edges annotated as 1.}
\initialize{
same as in Algorithm~\ref{alg:greedyAlgorithm}\;
}
\For{edge \(e\in E\)}{
\If{\(\max\limits_{c\in C(e)}\#\)refined edges of \(c\le4\)}{
\(E\gets E\setminus\{e\}\)\;
}
}
\(x^*\gets\mathop{\arg\min}\limits_x\sum\limits_{c\in C}\#\mathrm{hex}(\sum\limits_{i=1}^{12}2^{12-i}x_{e_c(i)})\) s.t. \(x_e\in\{0,1\}\) for edge \(e\in E,x_e=\hat x_e\) for edge \(e\notin E\)\;
\For{edge \(e\in E\)}{
\If{\(x^*_e=1\)}{
refine edge \(e\)\;
}
}
\Return{modified edges}\;
\end{algorithm2e}

The ILP then determines the statuses of all edges in \(E\) simultaneously. Each edge \(e\) carries a status variable \(x_e\): for \(e\in E\), \(x_e\in\{0,1\}\) is a Boolean variable, whereas for \(e\notin E\), \(x_e\) is a determined status \(\hat x_e\in\{0,1\}\). For each cluster \(c\), let \(e_c(1),\dots,e_c(12)\) enumerate its 12 edges in a fixed order, so that its state is the 12-bit decimal value \(r_c(x)=\sum\limits_{i=1}^{12}2^{12-i}x_{e_c(i)}\in\{0,\dots,4\,095\}\). The ILP formulation is \(\min\limits_x\sum\limits_{c\in C}\#\mathrm{hex}(r_c(x))\text{ s.t. }x_e\in\{0,1\}\text{ for }e\in E,x_e=\hat x_e\text{ for }e\notin E\), i.e., it minimizes the total number of elements over all completions of the undetermined edge statuses. The pseudocode is given in Algorithm~\ref{alg:ILPAlgorithm}. Because ILP is NP-hard, the worst runtime is \(O(2^{\lvert E\rvert})\), where \(\lvert E\rvert\) is the number of undetermined edges, although in practice the solvers terminate far faster.

\begin{Remark}
Two remarks are in order. First, the element reduction achieved by applying Algorithm~\ref{alg:greedyAlgorithm} as a preprocessing step to Algorithm~\ref{alg:ILPAlgorithm} is virtually identical to that of Algorithm~\ref{alg:ILPAlgorithm} alone, even when the threshold is set to six. Hence, the greedy step can be omitted when ILP is employed. Second, because \(\#\mathrm{hex}\) is a step function of the state value \(r_c\), the objective \(\min\limits_x\sum\limits_{c\in C}\#\mathrm{hex}(r_c(x))\) is nonlinear. It is linearized by enumerating, for each cluster, all candidate states and selecting one via auxiliary Boolean variables. For cluster \(c\) with \(n_c\) undetermined edges \(e_c(i_1),\dots,e_c(i_{n_c})\), the determined edges contribute the constant offset \(\omega_c=\sum\limits_{j\notin\{i_1,\dots,i_{n_c}\}}2^{12-j}\hat x_{e_c(j)}\). For each state \(s\in\{0,\dots,2^{n_c}-1\}\), with \(j\)-th bit \(s_j\), a Boolean selector \(y_{c,s}\) subject to \(\sum\limits_sy_{c,s}=1\) and \(x_{e_c(i_j)}=\sum\limits_{s:s_j=1}y_{c,s}\) for \(j=1,\dots,n_c\) is introduced, so that each cluster occupies exactly one state whose bits are tied to the edge variables (which, as a byproduct, forces every \(x_e\) to be binary). The objective then becomes \(\min\limits_{x,y}\sum\limits_{c\in C}\sum\limits_{s=0}^{2^{n_c}-1}c_{c,s}y_{c,s}\), where each coefficient \(c_{c,s}=\#\mathrm{hex}(\omega_c+\sum\limits_{j=1}^{n_c}2^{12-i_j}s_j)\) is a precomputed constant; the objective is thus linearized.
\end{Remark}

\section{Results}
\label{sec:results}

The source code is publicly available at (code is attached in the supplementary material and will release on github upon acceptance). The implementation reads an input triangular mesh and constructs an initial octree based on the shape diameter function (SDF) \cite{shapira2008consistent}: a cell is refined when its edge length exceeds \(0.5\) of the minimum SDF value among the triangular faces contained within the cell \cite{pitzalis2021generalized}. The octree is then balanced, the hanging nodes are removed, and the resulting mesh is exported as an all-hex background mesh.

\begin{figure*}[ht]
\centering
\includegraphics[width=\linewidth]{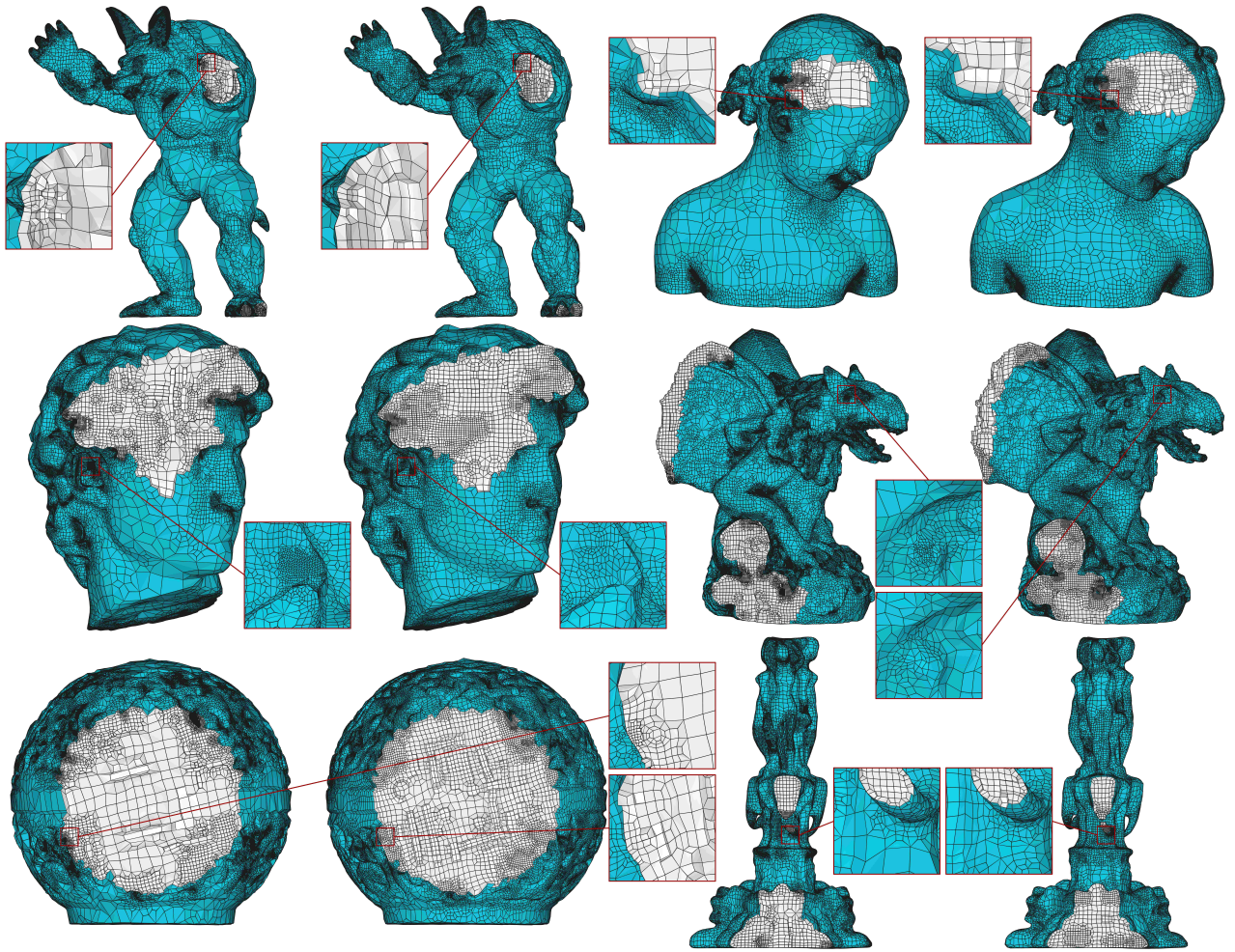}
\vspace{-8mm}
\caption{Comparison of \cite{tong2026element} (left) and the proposed method (right) on six randomly selected smooth models.}
\label{fig:resultsSmooth}
\end{figure*}

The only design decision involving a trade-off concerns the threshold in Algorithm~\ref{alg:ILPAlgorithm}: the ILP optimization skips an edge if the maximum number of already-refined edges among all clusters sharing that edge is no greater than the threshold. Generally, a smaller threshold leaves more edges to optimize, reducing the element count but increasing runtime. To select a proper value, program runtime is measured on the 202 benchmark models \cite{gao2019feature} for Algorithm~\ref{alg:ILPAlgorithm} with thresholds six, five, and four, for Algorithm~\ref{alg:greedyAlgorithm}, and for the baseline, which disables the element-reduction step. For each model, the runtime is normalized by that of the baseline, and the resulting ratios are averaged over the 202 models. The mean relative runtimes are 1.44 for the greedy algorithm and 2.64, 2.69, and 5.71 for the ILP algorithm with thresholds six, five, and four, respectively. Since a threshold of three is prohibitively slow on complex models, the threshold is set to four in Algorithm~\ref{alg:ILPAlgorithm}.

The two-refinement scheme is compared with the three-refinement scheme \cite{tong2026element} on the same benchmark. Following the pipeline of \cite{tong2026element}, both approaches generate hex meshes free of hanging nodes. For a fair comparison, the total element count of the three-refinement mesh is matched to that of the two-refinement greedy method by adjusting the edge-length threshold multiplier starting from \(0.5\). The multiplier is tuned so that the three-refinement mesh contains slightly more hex elements. Following \cite{pitzalis2021generalized}, each cell is classified according to the signed distance values at its eight vertices and retained only if their sum is negative; surface elements that would otherwise become low-quality after optimization are removed \cite{tong2024hybridoctree_hex}; and mesh quality and surface fitting are optimized for 500 iterations \cite{protais2026versatile}. Finally, the HR and min SJ of the two optimized meshes are computed.

\begin{figure*}[ht]
\centering
\includegraphics[width=\linewidth]{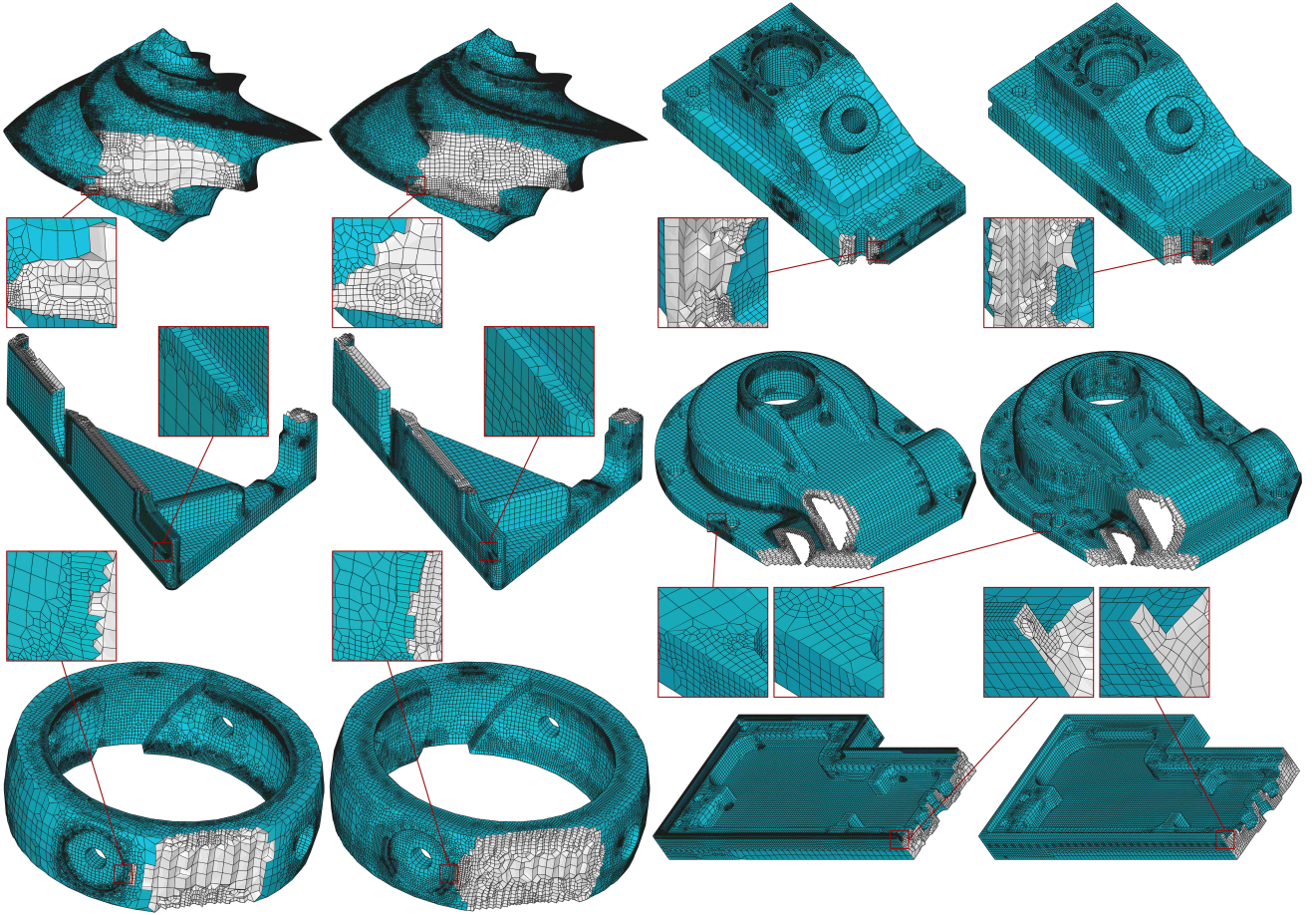}
\vspace{-8mm}
\caption{Comparison of \cite{tong2026element} (left) and the proposed method (right) on six randomly selected CAD models.}
\label{fig:resultsCAD}
\end{figure*}

\Cref{fig:resultsSmooth,fig:resultsCAD} compare \cite{tong2026element} with the proposed method on six smooth and six CAD models, respectively, randomly selected from the 202 test models; red boxes zoom into details. Conceptually, two-refinement relates to three-refinement much as binary search relates to ternary search. As shown, three-refinement tends to produce steeper transitions, whereas two-refinement yields smoother ones, consistent with reports in \cite{tong2026element}.

\begin{table}[ht]
\centering
\caption{Comparison of five proposed methods against \cite{tong2026element} over 202 models. Each cell gives the mean of per-model ratios (method value divided by \cite{tong2026element} value). A ratio below/above 1 means the method outperforms/underperforms \cite{tong2026element} for desired \(\downarrow\)/\(\uparrow\) metrics. Bold values are best values.}
\label{tab:comparison}
\setlength{\tabcolsep}{4pt}
\begin{tabular}{c|ccccc}
metrics&baseline&greedy&ILP-6&ILP-5&ILP-4\\
\hline
\#vert\(\downarrow\)&\(0.958\)&\(0.941\)&\(0.922\)&\(0.921\)&\(\mathbf{0.917}\)\\
\#hex\(\downarrow\)&\(0.944\)&\(0.929\)&\(0.908\)&\(0.907\)&\(\mathbf{0.902}\)\\
HR\(\downarrow\)&\(1.01\)&\(1.00\)&\(\mathbf{0.987}\)&\(0.993\)&\(1.03\)\\
min SJ\(+1\)\(\uparrow\)&\(1.056\)&\(1.056\)&\(1.060\)&\(1.060\)&\(\mathbf{1.062}\)\\
\end{tabular}
\end{table}

Table~\ref{tab:comparison} compares five two-refinement variants, the baseline, greedy, and ILP with thresholds six, five, and four, against three-refinement \cite{tong2026element} on 202 models. The four metrics from top to bottom are total vertices, total hex elements, HR after optimization \cite{protais2026versatile}, and min SJ\(+1\); since min SJ \(\in[-1,1]\), it is shifted by \(+1\) to keep the ratio denominator nonzero. From left to right, both resulting vertex and hex counts decrease. All five variants achieve higher min SJ than \cite{tong2026element} while producing slightly fewer elements, because two-refinement yields smoother transitions and more regular singularities than three-refinement. For HR, however, two-refinement shows no clear advantage: since three-refinement is not subject to the pairing condition, it can focus refinement on more detailed regions, saving elements elsewhere. This agrees with the comparison in \cite{tong2026element} against \cite{pitzalis2021generalized}.

\begin{figure*}[ht]
\centering
\includegraphics[width=\linewidth]{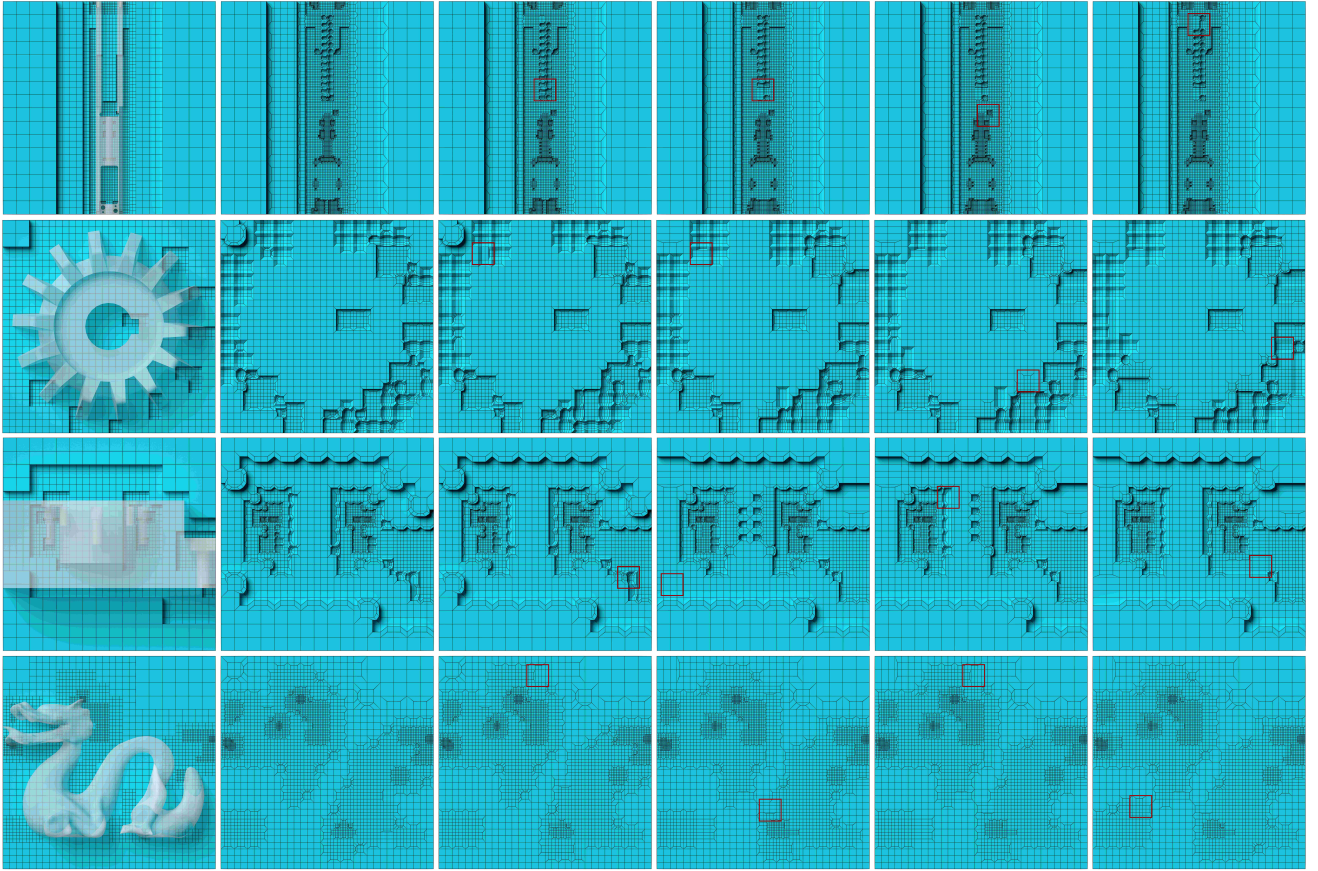}
\vspace{-8mm}
\caption{Two-refinement hex meshes of four randomly selected models. From left to right: the input geometry with the initial octree (after moderate balancing and octree pairing), then meshes from the baseline, the greedy method, and the ILP with thresholds of six, five, and four. Red boxes highlight local changes on the cross-section.}
\label{fig:elementReduction}
\end{figure*}

The element-reduction stage is shown on four randomly selected models in \Cref{fig:elementReduction}. Each row shows, from left to right, the initial octree after moderate balancing and octree pairing and the meshes produced by baseline, greedy, and ILP (thresholds of six, five, and four) methods. From left to right, runtime grows while element count drops. The moderate balancing and octree pairing yield a larger initial octree than in \cite{pitzalis2021generalized}; in return, all hex faces are planar. Thus, the proposed method is currently the only two-refinement method compatible with quality-guaranteed hex meshing \cite{tong2025mchex}.

\section{Conclusion and future work}
\label{sec:conclusionAndFutureWork}

This paper presents a primal two-refinement algorithm for hanging-node removal. Unlike existing primal two-refinement methods \cite{schneiders2000octree,ebeida2011isotropic,huang2013incorporating,owen2017template}, which enforce a strong balancing condition and over-refine the grid, the proposed method enforces a more relaxed moderate balancing condition. It generalizes the edge-based three-refinement templates of \cite{tong2026element} to two-refinement: 32 special patterns and five fundamental patterns together resolve all 144 symmetry-reduced configurations. The fundamental patterns project the internal cluster of a local refinement onto the \(1100\) faces and decouple the remaining small clusters, so that all configurations can be handled; the resulting element count is further reduced either by a fast greedy algorithm or by an ILP algorithm that trades additional runtime for further savings. Experiments on the 202 models of the \cite{gao2019feature} dataset demonstrate that the proposed scheme attains a similar HR and a higher min SJ, while the meshes generated by three-refinement contain slightly more elements. This is the first two-refinement method in which every hex element has planar faces; the resulting meshes therefore satisfy the planarity requirement of MCHex \cite{tong2025mchex}, a quality-guaranteed reconstruction framework, and can serve as a drop-in replacement for the three-refinement templates it currently employs.

The current method has several limitations that point to directions for future research. First, the proposed templates assume that all cells at each octree level can be grouped into clusters without overlap or omission, an assumption that may not hold under general pairing. This does not, however, preclude adapting the template idea to that setting, and generalizing the template set to general pairing is a natural next step. Extending the framework to weak balancing would trade additional singularities for a further reduction in element count, an incremental improvement, but a worthwhile one. Second, the ILP is currently too slow to solve without pruning; therefore, simplifying the formulation can enable faster optimization without compromising element reduction performance. Finally, machine learning, LLMs, and AI agents can reduce the heuristic steps involved in the meshing pipeline \cite{tong2023srl,yu2025dl,yu2026ddpm,yu2026eddpm,owen2026survey}. One direction is to employ reinforcement learning to search for more compact two-refinement transition templates, with the aim of generating meshes with fewer elements.

\section*{Acknowledgments}
This project was supported in part by the US National Science Foundation grants CMMI-1953323 and CBET-2332084.

\bibliographystyle{siam}
\bibliography{ltexpprt_references}
\end{document}